\documentclass{article}

\usepackage{spconf}

\usepackage{amsmath,amssymb,amsfonts,amsthm,mathtools}
\usepackage{optidef}
\usepackage{algorithm,algorithmic}
\usepackage{graphicx}
\usepackage{cite,hyperref}
\usepackage[capitalise]{cleveref}
\newcommand{\letterref}[2]{\namecref{#1}~\hyperref[#1]{\ref*{#1}#2}}

\usepackage{pgfplots}
\usepgfplotslibrary{groupplots}
\usepgfplotslibrary{colormaps}
\pgfplotsset{compat=1.16}
\pgfplotscreateplotcyclelist{mycyclelist}{%
  {red, mark=o, mark options={solid}, solid, thick},
  {blue, mark=x, mark options={solid}, dashed, thick},
  {orange, mark=*, mark options={solid}, dotted, thick},
  {violet, mark=o, mark options={solid}, dashed, thick},
  {red, mark=*, mark options={solid}, dashed, thick},
  {black, mark=x, mark options={solid}, solid, thick},
}
\pgfplotsset{cycle list name=mycyclelist}

\usepackage{color}

\usepackage[normalem]{ulem}

\newcommand{\argmin}{\operatornamewithlimits{argmin}}

\DeclareMathOperator{\tr}{tr}

\DeclareMathOperator{\diag}{diag}

\newcommand{\opnorm}[1]{\ensuremath{{\left\vert\kern-0.25ex\left\vert\kern-0.25ex\left\vert #1 \right\vert\kern-0.25ex\right\vert\kern-0.25ex\right\vert}}}

\newcommand{\R}{\ensuremath{{\mathbb R}}}
\newcommand{\Rn}{\ensuremath{{\mathbb R}^n}}
\newcommand{\Snn}{\ensuremath{{\mathbb S}^{n\times n}}}

\newcommand{\Nodes}{\ensuremath{{\mathcal V}}}
\newcommand{\Edges}{\ensuremath{{\mathcal E}}}
\newcommand{\Graph}{\ensuremath{{\mathcal G}}}

\newcommand{\W}{\ensuremath{{\mathcal W}}}
\newcommand{\Filt}{\ensuremath{{\mathcal H}}}

\newcommand{\Cy}{\ensuremath{{\mathbf C}_{\mathbf y}}}
\newcommand{\Cyhat}{\ensuremath{\widehat{\mathbf C}_{\mathbf y}}}

\newcommand{\I}{\ensuremath{{\mathbf I}}}
\newcommand{\A}{\ensuremath{{\mathbf A}}}

\renewcommand{\L}{\ensuremath{{\mathbf L}}}
\newcommand{\D}{\ensuremath{{\mathbf D}}}

\newcommand{\blambda}{\ensuremath{{\boldsymbol \lambda}}}

\newcommand{\x}{\ensuremath{{\mathbf x}}}
\renewcommand{\v}{\ensuremath{{\mathbf v}}}
\newcommand{\y}{\ensuremath{{\mathbf y}}}
\newcommand{\w}{\ensuremath{{\mathbf w}}}
\renewcommand{\d}{\ensuremath{{\mathbf d}}}

\newcommand{\drv}{\ensuremath{\mathrm d}}

\newcommand{\iid}{\textit{i.i.d.}}
\newcommand{\eg}{\textit{e.g.}}
\newcommand{\ie}{\textit{i.e.}}

\newcommand{\ER}{{Erd\H{o}s-R\'enyi}}

\newtheorem{problem}{Problem}
\newtheorem{theorem}{Theorem}

\newtheoremstyle{myremark}
{\topsep} 
{\topsep} 
{\normalfont} 
{} 
{\bfseries} 
{.} 
{5pt plus 1pt minus 1pt} 
{\thmname{#1}\thmnumber{ #2}\thmnote{ (#3)}} 

\theoremstyle{myremark}
\newtheorem{remark}{Remark}

\newcommand{\parheading}[1]{\vspace{1mm}\noindent\textbf{#1}}
\newcommand{\subparheading}[1]{\vspace{1mm}\noindent\textit{#1}}

\title{Network topology inference with graphon spectral penalties}
\name{T. Mitchell Roddenberry, Madeline Navarro, and Santiago Segarra
  \thanks{This work was partially supported by NSF under award CCF-2008555.
    Emails: \href{mailto:mitch@rice.edu}{mitch@rice.edu}, \href{mailto:madeline.navarro@rice.edu}{madeline.navarro@rice.edu}, \href{mailto:segarra@rice.edu}{segarra@rice.edu}}}
\address{Rice University, Electrical and Computer Engineering, Houston, TX}

\begin{document}
\ninept
%
\maketitle
\begin{abstract}
  We consider the problem of inferring the unobserved edges of a graph from data supported on its nodes.
  In line with existing approaches, we propose a convex program for recovering a graph Laplacian that is approximately diagonalizable by a set of eigenvectors obtained from the second-order moment of the observed data.
  Unlike existing work, we incorporate prior knowledge about the distribution from where the underlying graph was drawn.
  In particular, we consider the case where the graph was drawn from a graphon model, and we supplement our convex optimization problem with a provably-valid regularizer on the spectrum of the graph to be recovered.
  We present the cases where the graphon model is assumed to be known and the more practical setting where the relevant features of the model are inferred from auxiliary network observations.
  Numerical experiments on synthetic and real-world data illustrate the advantage of leveraging the proposed graphon prior, even when the prior is imperfect.
\end{abstract}

\section{Introduction}\label{sec:intro}

Networks have emerged as an effective tool for analyzing and summarizing complex systems in science and engineering~\cite{strogatz2001, jackson2010, newman2010}.
The abstraction of these systems as graphs enables one to examine local structures of influence among agents via direct observation of the edge structure, as well as global behavior via the spectral features of algebraic representations of these graphs.
The algebraic properties of networks are used to succinctly describe their behavior, such as the centrality structure~\cite{pagerank1999}, community structure~\cite{von2007tutorial}, or conductance~\cite{levin2017markov}.
In some settings, a network is known to the user or can be simply constructed, \eg{} a rail transportation network, where each station is a node connected by train routes modeled by edges.
However, the network structure must often be inferred from the system behavior itself.
For instance, the system of functional relationships between regions of the brain may not be known directly, but needs to be inferred from observed neural activity.
This is an example of the well-established problem of \emph{network topology inference}.

In this work, we seek to \emph{infer the structure of the graph Laplacian} from so-called \emph{spectral templates}~\cite{segarra2017network}, \ie{}, by first estimating the eigenvectors of the Laplacian matrix from the observed data and then selecting the eigenvalues that yield a graph Laplacian with desirable features.
Work in this direction has studied how the structure of a graph shapes data supported on it and how to invert that process for network topology inference, but has neglected the structure endowed by common statistical models of real-world graphs.
For instance, wireless sensor networks are well-modeled as geometric random graphs~\cite{kenniche2010random}, and social networks often follow power-law degree distributions~\cite{muchnik2013origins}.
In this work, we aim to imbue the network topology inference pipeline with this type of domain knowledge.
As a flexible random graph model, we assume the inferred graph is drawn from a graphon, a non-parametric model for large random graphs~\cite{borgs2017graphons,lovasz2012large}.
Graphons have proven to be an effective model for understanding large graphs, acting as a limit object for the graphs they generate~\cite{morency2017signal,lee2017unifying,gau2020gcol,avella2020centrality}.
In this work, we leverage convergence results on graphs generated by graphons~\cite{vizuete2020laplacian} to regularize the spectral structure of the inferred graph Laplacian.

\parheading{Related work.}
The problem of network topology inference is well-studied from a variety of perspectives, each incorporating different assumptions on the data model and imposing characteristics on the graph structure.
The statistical perspective represents each node as a random variable, where the graph structure between them captures conditional dependencies~\cite{koller2009probabilistic}.
Network inference then seeks to find a sparse inverse covariance (precision) matrix~\cite{friedman2008sparse,lake2010discovering}.
Indeed, the notion of incorporating constraints on the combinatorial structure of the inferred graph in the context of probabilistic graphical models was studied in~\cite{kumar2020unified}.
This differs from our work here, where the graph is assumed to be drawn from a distribution exhibiting certain combinatorial properties in a statistical sense, but lacking hard combinatorial constraints.

On the other hand, modeling network processes as the description of a physical system yields another set of methodologies.
In these works, topology inference is performed under models of systems related to information diffusion, random walks, and related network processes~\cite{gomez2012inferring,shahrampour2014topology}.
These techniques are closely related to those in the graph signal processing literature~\cite{mateos2019connecting}, where the observed data is modeled as a set of graph signals, which are assumed to be smooth on the underlying graph~\cite{dong2016learning,kalofolias2016learn}, or the output of a graph filter applied to excitation noise~\cite{segarra2017network,pasdeloup2017characterization}, perhaps modeling heat diffusion~\cite{thanou2017learning} or consensus dynamics~\cite{shahrampour2013reconstruction,zhu2020network}.
Although our observation model is in line with this body of work, the novelty lies in the incorporation of prior statistical knowledge of the graph to be recovered in the form of a graphon model.

\parheading{Contribution.}
The contributions of this paper are twofold:\\
i) We propose an efficient convex optimization problem for network topology inference with a novel spectral shrinkage penalty that leverages the relationship between the \emph{spectrum of the Laplacian} and the \emph{degree distribution of graphs drawn from graphons}.\\
ii) We then demonstrate that this approach has practical utility beyond the case where the underlying graphon is known, by substituting the limiting (graphon) degree distribution with empirically obtained degree statistics.

\section{Preliminaries}\label{sec:bg}


\parheading{Notation.}
The notation $[n]$ refers to the set of positive integers $\{1,2,\ldots,n\}$.
We refer to matrices using bold uppercase letters, \eg{} $\mathbf{A,B,C}$, and to (column) vectors with bold lowercase letters, \eg{} $\mathbf{v,w,x}$.
Entries of matrix $\A$ are indicated by $A_{ij}$ while those of vector $\x$ are denoted by $x_i$.
For clarity, we alternatively use the notation $[\x]_i=x_i$.
The $\ell_2$-norm of a vector is denoted by $\|\cdot\|_2$.
The set of symmetric $n\times n$ matrices is denoted by $\Snn$.
We use the notation $\|\cdot\|_{1,1}$ to denote the element-wise $\ell_1$-norm of a matrix, and $\|\cdot\|_F$ for the Frobenius norm.
For a linear integral operator $W$ on $L^2[0,1]$, its \emph{operator norm} is specified by $\opnorm{W}$:
\begin{equation}\label{eq:opnorm}
  \opnorm{W}=
  \sup_{\substack{f\in L^2[0,1]\\ \|f\|_2=1}}
  \int_{[0,1]^2} W\left(x,y\right)f\left(x\right) \,\drv{x}\,\drv{y}.
\end{equation}

\parheading{Graphs, graph signals, and graph filters.}
A \emph{graph} is a finite set of nodes, coupled with a set of edges connecting those nodes.
That is, for a set of $n$ nodes denoted by $\Nodes$, the set of edges $\Edges\subseteq\Nodes\times\Nodes$ forms the graph $\Graph=(\Nodes,\Edges)$.
Typically, we endow the set of nodes with arbitrarily ordered integer labels, saying that $\Nodes=[n]$.
This representation allows us to represent the graph with the \emph{adjacency matrix} $\A\in\R^{n\times n}$, where $A_{ij}=1$ if $(i,j)\in\Edges$, taking value $0$ otherwise.
A graph is \emph{undirected} if for any two nodes $i,j$, $(i,j)\in\Edges$ implies that $(j,i)\in\Edges$: that is, $\Edges$ is composed of unordered tuples of nodes.
Under this condition, the adjacency matrix is symmetric.
In this setting, we define the \emph{Laplacian matrix} as $\L=\D-\A$,
where $\D=\diag(\A{\mathbf 1})$ is the diagonal matrix of node degrees.

We model data on the nodes of a graph as a \emph{graph signal}.
A graph signal is a real-valued function on the nodes of a graph $x:\Nodes\to\R$.
If we endow the set of nodes with a labeling of positive integers so that $\Nodes=[n]$, then a graph signal $x$ has a natural representation $\x$ in $\Rn$, where $[\x]_i=x(i)$ for each $i\in[n]$.

With graph signals represented as vectors in $\Rn$, we now define \emph{graph filters} as linear maps between them.
Assuming that $\L$ has the eigenvalue decomposition $\L = \sum_i\lambda_i\v_i\v_i^\top$, a graph filter is a real polynomial of $\L$, \eg{} for coefficients $\alpha_k$,
\begin{equation}\label{eq:graph-filter}
\Filt\left(\L\right)=\sum_{k=0}^T\alpha_k\L^k=\sum_{k=0}^Th\left(\lambda_i\right)\v_i\v_i^\top,
\end{equation}
where $h:\R\to\R$ is the extension of $\Filt$ to the real numbers.\footnote{Graph filters can be defined in terms of generic graph shift operators~\cite{segarra2017optimal}, but here we focus on those defined based on $\L$.}
Notice that graph filters preserve the eigenvectors of the underlying Laplacian, while distorting the eigenvalues.

\parheading{Graphons as random graph models.}
A \emph{graphon} is a symmetric, measurable function from the unit square $[0,1]^2$ to the unit interval $[0,1]$, with the set of all such functions denoted by $\W$~\cite{borgs2017graphons}.
For a graphon $W\in\W$, we sample a graph of size $n$ by first drawing \iid{} uniform random variables $\{\zeta_i\}_{i=1}^n\sim{\mathcal U}[0,1]$.
Then, the adjacency matrix of our drawn graph is determined by setting $A_{ij}=A_{ji}=1$ with probability $W(\zeta_i,\zeta_j)$, and $A_{ij}=A_{ji}=0$ otherwise.
A graphon determines a probability distribution over graphs of fixed size in this way.
For instance, the constant graphon $W(x,y)=p$ for all $x,y\in[0,1]$ yields \ER{} graphs of size $n$ and edge density $p$.
From this sampling method, one can see that any measure-preserving bijection $\tau:[0,1]\to[0,1]$ yields an equivalence relation between graphons.
That is, if for all $x,y\in[0,1]$ we say that $W^\tau(x,y)=W(\tau(x),\tau(y))$, then $W^\tau\sim W$.
Additionally, if two graphons $W^0,W^1$ are equal almost everywhere, we also say that $W^0\sim W^1$.

\section{Network topology inference}\label{sec:nti}

In order to formally define our problem, we need to assume a model on how the underlying graph structure shapes the data on its nodes.
Without such a relationship, inferring the graph from the observed graph signals would be a hopeless endeavor.
To achieve this, we leverage the versatility of graph filters in representing network processes~\cite{segarra2017optimal} and model the observed signals as the outputs of graph filters.
For simplicity we focus on a class of network processes corresponding to discrete-time consensus dynamics, a well-studied model for network topology inference~\cite{shahrampour2013reconstruction,zhu2020network}.
More precisely, we model our observed graph signals $\y\in\Rn$ as being generated by
\begin{equation}\label{eq:graph-signal-model}
\y = \Filt\left(\L\right) \w = \prod_{k=1}^T\left(\I-\alpha_k\L\right)\w,
\end{equation}
with unknown coefficients $0\leq\alpha_k\leq\lambda_{max}(\L)$ and $\w$ being an unknown random vector drawn from a zero-mean distribution with identity covariance.
With this notation in place, we can formally state our problem.
\begin{problem}\label{prob:nti}
  Consider an unknown graph $\Graph$ drawn from a known graphon $W$. Given a collection of graph signals $\{\y^{\ell}\}_{\ell=1}^m$ defined in $\Graph$ following the model in~\eqref{eq:graph-signal-model}, estimate the Laplacian $\L$ of $\Graph$.
\end{problem}
\noindent
Given perfect knowledge of a graph filter $\Filt\left(\L\right)$, it follows from~\eqref{eq:graph-filter} that the eigenvectors of $\L$ can be readily obtained from an eigendecomposition of $\Filt\left(\L\right)$.
Moreover, even if the filter is unknown, we can estimate $\{\v_i\}_{i=1}^n$ from the observations in $\{\y^{\ell}\}_{\ell=1}^m$ as in previous works~\cite{segarra2017network, mateos2019connecting}.
In a nutshell, a simple computation of the covariance $\Cy$ of the signals in~\eqref{eq:graph-signal-model} reveals that
\begin{equation}\label{eq:covariance}
\Cy = \Filt^2\left(\L\right) = \sum_{k=0}^T h^2\left(\lambda_i\right)\v_i\v_i^\top,
\end{equation}
so that the covariance matrix of $\y$ shares a set of eigenvectors with $\L$. 
Thus, from the observed signals $\{\y^{\ell}\}_{\ell=1}^m$ we may estimate $\Cy$ via a sample covariance estimator $\Cyhat$ and obtain approximate eigenvectors (or spectral templates) $\{\widehat{\v}_i\}_{i=1}^n$.
Note that in obtaining these estimated spectral templates we have not used the knowledge of the specific form of the network process in~\eqref{eq:graph-signal-model} nor any knowledge of the graphon $W$. 
In~\cref{sec:nti:approach}, we present a convex optimization problem that takes as inputs $\{\widehat{\v}_i\}_{i=1}^n$ and these additional sources of information to provide an estimate of $\L$, as desired in~\cref{prob:nti}.

\begin{remark}[Alternative problem formulations]\label{R:alternative_problems}
  Although we focus on~\cref{prob:nti} as stated, several variations and relaxations can be tackled by incorporating the methodology proposed here with existing approaches.
  For instance, knowledge of the coefficients $\alpha_k$ in~\eqref{eq:graph-signal-model} can be further exploited as in~\cite{zhu2020network}, and the identity covariance of $\w$ can be relaxed as in~\cite{shafipour2017network,shafipour2018identifying}.
  Moreover, since the signals $\{\y^{\ell}\}_{\ell=1}^m$ are used to obtain the estimated eigenvectors $\{\widehat{\v}_i\}_{i=1}^n$, the optimization method proposed in the next section can be directly applied given any other source of approximated spectral templates~\cite{segarra2017network}.
  Lastly, we might consider the case where the graphon $W$ is also unknown.
  This empirically-relevant variant is discussed toward the end of~\cref{sec:nti:approach} and illustrated in~\cref{sec:exp}.
\end{remark}

\subsection{Inferring the Laplacian with graphon penalties}\label{sec:nti:approach}

We now consider how knowledge of the graphon $W$ from which a graph was drawn can inform our inferred Laplacian spectrum.
The view of graphons as generative models for dense, simple graphs yields probabilistic bounds relating properties of graphs drawn from a graphon and the analogous property of the graphon itself.
For instance, it is stated in~\cite[Corollary 10.4]{lovasz2012large} that the density of a motif in a sufficiently large graph concentrates around the density of that motif in the graphon from which it was drawn, \eg{} the homomorphism density of triangles in a graph:
\begin{equation}\label{eq:graphon:tri-density}
  \frac{1}{n^3}\tr\left(\A^3\right)\overset{n}{\to}\int_{[0,1]^3}W\left(x,y\right)W\left(y,z\right)W\left(z,x\right)\ \drv{x}\,\drv{y}\,\drv{z}.
\end{equation}
In particular, the following result characterizes the \emph{spectrum of the Laplacian matrix} of large graphs drawn from a graphon.
\begin{theorem}[From \cite{vizuete2020laplacian}]\label{thm:graphon-concentration}
  Let $W$ be a piecewise-Lipschitz graphon such that $W(x_1,y)\leq W(x_2,y)$ when $x_1\leq x_2$, and the infimum of $W$ over $[0,1]^2$ is bounded away from $0$.
  Then, for a graph $\Graph$ of size $n$ sampled from $W$, where $0=\lambda_1\leq\ldots\leq\lambda_n$ are the eigenvalues of the Laplacian matrix of \Graph, with probability at least $1-3\nu$,
  \begin{equation}
    \label{eq:graphon:laplace-spectrum}
    \|\mu(x)-d(x)\|_2\leq C_0 + \sqrt[4]{\frac{2}{n}}\sqrt{\opnorm{W} + C_1} + C_1,
  \end{equation}
  where $\mu(x)=\lambda_{\lfloor{nx}\rfloor+1}/n$ for all $x\in[0,1)$, $\mu(1)=\lambda_n/n$, $d(x)=\int_0^1W(x,y)\drv{y}$, and $C_0,C_1$ are dependent on $\nu$, the Lipschitz smoothness of $W$, and the infimum of $W$.
\end{theorem}
\noindent 
Notice that $\mu(x)$ is a piecewise constant function given by the eigenvalues of the Laplacian and the degree function $d(x)$ is the analogous of a degree sequence defined directly on the continuous graphon~\cite{avella2020centrality}.
Informally, \cref{thm:graphon-concentration} states that the ordered Laplacian eigenvalues concentrate near the ordered degree function as $n$ grows, especially for smooth graphons whose values are bounded far away from zero.
It is key to notice that this result does not depend on knowing the particular degrees of the nodes in the graph nor their latent positions $\zeta_i\in[0,1]$ in the sampling procedure (cf. Section~\ref{sec:bg}).


With~\cref{thm:graphon-concentration} in hand, we propose a method for the recovery of the graph Laplacian, with tolerance for imperfections in the spectral templates brought about by noise or finite sampling effects.
We combine an $\ell_1$-penalty on the inferred Laplacian matrix to promote edge-sparsity~\cite{segarra2017network,zhu2020network}, as well as a spectral shrinkage penalty towards a prior degree function.
One key property of the class of filters in~\eqref{eq:graph-signal-model} is that the extension of $\Filt$ to the real numbers $h:\R\to\R$ is \emph{monotonically decreasing} and \emph{non-negative} in the spectrum of $\L$, so that it preserves, in reverse, the ordering of the eigenvalues of $\L$, even after being squared as in~\eqref{eq:covariance}.
This is captured in the following convex program, where the spectral templates $\{\widehat{\v}_i\}_{i=1}^n$ are used as an approximate ordered eigenbasis for the inferred matrix,
\begin{align}\label{eq:recovery-program}
  \L^*,\blambda^* & = \argmin_{\L,\blambda} \|\L\|_{1,1} + \beta\|\mu(x)-d(x)\|_2^2 \\
  \mathrm{s.t.}\quad & \rho\left(\L,\sum_{i=1}^n\lambda_i\widehat{\v}_i\widehat{\v}_i^\top\right)\leq\epsilon, \,\,\, \L\in{\mathcal L}, \nonumber \\
                  & \mu(x)=\lambda_{\lfloor nx\rfloor+1}/n \,\quad\text{for all }x\in[0,1), \nonumber \\
                  & \mu(1)=\lambda_n/n, \nonumber \\
                  & \lambda_i\leq\lambda_{i+1+\eta} \qquad\quad\enskip\text{ for all }i\in[n-1-\eta], \nonumber 
\end{align}
where $\mathcal L$ indicates the set of valid graph Laplacians, the function $\rho:\Snn\times\Snn\to\R^+$ is a distance function on symmetric matrices, and $\epsilon$ dictates how far the eigenvectors of the inferred graph Laplacian can be from the approximated spectral templates.
The parameter $\eta$ allows for slack in the ordering of the eigenvalues, due to spectral perturbation in the estimation of the spectral templates.
Finally, $\mu(x)$ and $d(x)$ are the same as in~\cref{thm:graphon-concentration}.

The program~\eqref{eq:recovery-program} jointly recovers a spectrum $\blambda^*$ and a valid Laplacian that is close to $\blambda^*$ in the eigenbasis $\{\widehat{\v}_i\}_{i=1}^n$.
The spectrum is encouraged to be close to the degree function of the graphon based on~\cref{thm:graphon-concentration}, and sparsity is promoted in the Laplacian.
We emphasize that shrinking the spectrum of the Laplacian towards the graphon degree function \emph{does not} require knowledge of the degrees of particular nodes.
That is, since the spectral shrinkage penalty operates only on the eigenvalues of $\L$, it is blind to permutations of the nodes.

\parheading{Inference when the graphon is unknown.}
In practice, it may be unrealistic to assume that the graphon $W$ is known.
However, the rich statistical structures of this modeling assumption can still be leveraged for the application of~\eqref{eq:recovery-program}.
Indeed, via~\cref{thm:graphon-concentration} we have distilled the signature of the graphon $W$ down to its degree function $d(x)$.
Therefore, if we can obtain an estimate of $d(x)$ from auxiliary information, then~\eqref{eq:recovery-program} can still be applied.
If a sufficiently large graph $\Graph=(\Nodes,\Edges)$ is drawn from an unknown graphon $W$, then its degree distribution will be similar to that of $W$.
In particular, suppose that a uniformly-sampled random subgraph of $\Graph$ is observed, where a subset of the nodes $\Nodes_0\subset\Nodes$ induces the graph $\Graph_0=(\Nodes_0,\Edges\cap\Nodes_0\times\Nodes_0)$.
It is clear that $\Graph_0$ is also distributed according to the sampling procedure described in~\cref{sec:bg} applied to $W$.
Therefore, the expected normalized degree distribution of $\Graph_0$ is approximately the graphon degree function $d(x)$, following naturally by considering the homomorphism density of a star graph as a way to query the degree distribution of a graph~\cite[Example~2.2]{borgs2006counting}.

With this in mind, we propose a modification to~\eqref{eq:recovery-program}, where a uniformly-sampled induced subgraph $\Graph_0$ on $n_0$ nodes of $\Graph$ is observed, with ordered vector of degrees $\d_0=\mathrm{sort}(\A_0{\mathbf 1})$.
In this case, the natural approximate degree function is $\widehat{d}(x)=[\d_0]_{\lfloor n_0x \rfloor+1}/n_0$ for $0\leq x< 1$, and $\widehat{d}(1)=[\d_0]_{n_0}$, which we use to then solve~\eqref{eq:recovery-program} with $d(x)$ replaced by $\widehat{d}(x)$.
Finally, even if the observation of a subgraph $\Graph_0$ is infeasible, the assumption of a graphon distribution can be used to incorporate information from similar graphs potentially supported on a different set of nodes.
For instance, when inferring the topology of a particular brain network, using the degree distribution of another, better-mapped brain would be a reasonable approximation of the assumed graphon degree function $W$ for use in~\eqref{eq:recovery-program}.

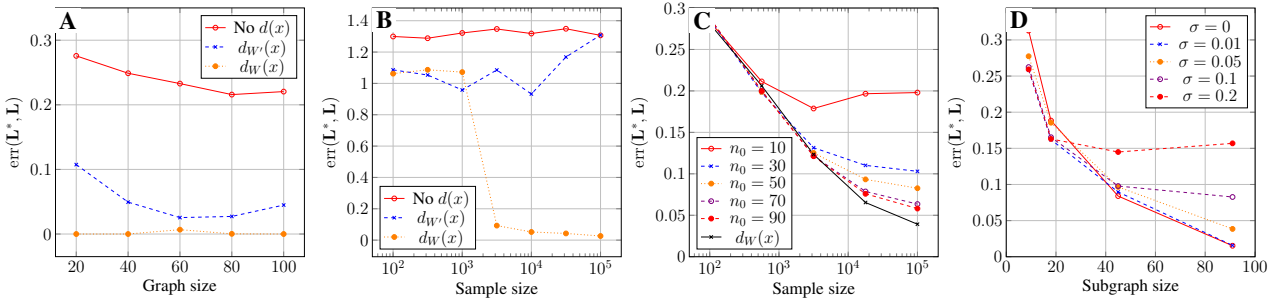
\begin{figure*}[t!]
  \centering
  \resizebox{0.95\textwidth}{!}{\begin{tikzpicture}

  \begin{groupplot}[
    group style={
      group name=myplots,
      group size=4 by 1,
      horizontal sep=2cm},
    legend style={
      font=\Large,
      fill opacity=1,
      text opacity=1,},
    every axis label/.append style={font=\Large},
    xmajorgrids,
    ymajorgrids,
    %
    tick align=inside,
    tick pos=both,
    xticklabel style={
      font=\Large,
      /pgf/number format/fixed,
      /pgf/number format/precision=2,
    },
    yticklabel style={
      font=\Large,
      /pgf/number format/fixed,
      /pgf/number format/precision=2,
    },
    width=0.5\textwidth,
    height=0.5\textwidth
    ]

    \nextgroupplot[
    xmode=normal,
    ymode=normal,
    ymax=0.35,
    legend pos=north east,
    xlabel={Graph size},
    ylabel={$\mathrm{err}(\L^*,\L)$},
    ]
    \addplot table[x=s, y=err] {./output/graphsize/frob_error_size_no_prior.csv};
    \addlegendentry{No $d(x)$}
    \addplot table[x=s, y=err] {./output/graphsize/frob_error_size_alt_prior.csv};
    \addlegendentry{$d_{W'}(x)$}
    \addplot table[x=s, y=err] {./output/graphsize/frob_error_size.csv};
    \addlegendentry{$d_W(x)$}

    \nextgroupplot[
    xmode=log,
    ymode=normal,
    legend pos=south west,
    xlabel={Sample size},
    ylabel={$\mathrm{err}(\L^*,\L)$},
    ]
    \addplot table[x=n, y=err] {./output/noisy/frob_error_noisy_no_prior.csv};
    \addlegendentry{No $d(x)$}
    \addplot table[x=n, y=err] {./output/noisy/frob_error_noisy_alt_prior.csv};
    \addlegendentry{$d_{W'}(x)$}
    \addplot table[x=n, y=err] {./output/noisy/frob_error_noisy.csv};
    \addlegendentry{$d_W(x)$}

    \nextgroupplot[
    xmode=log,
    ymode=normal,
    ymin=0, ymax=0.3,
    legend pos=south west,
    xlabel={Sample size},
    ylabel={$\mathrm{err}(\L^*,\L)$},
    ]
    \addplot table[x=m, y=err] {./output/subsample/n100_10.csv};
    \addlegendentry{$n_0=10$}
    \addplot table[x=m, y=err] {./output/subsample/n100_30.csv};
    \addlegendentry{$n_0=30$}
    \addplot table[x=m, y=err] {./output/subsample/n100_50.csv};
    \addlegendentry{$n_0=50$}
    \addplot table[x=m, y=err] {./output/subsample/n100_70.csv};
    \addlegendentry{$n_0=70$}
    \addplot table[x=m, y=err] {./output/subsample/n100_90.csv};
    \addlegendentry{$n_0=90$}
    \addplot table[x=m, y=err] {./output/subsample/n100_W.csv};
    \addlegendentry{$d_W(x)$}
  
    \nextgroupplot[
    xmode=normal,
    ymode=normal,
    xmin=0,
    ymin=0,
    legend pos=north east,
    xlabel={Subgraph size},
    ylabel={$\mathrm{err}(\L^*,\L)$},
    ]
    \addplot table[x=n0, y=ave] {./output/subsample/macaque_0_00.csv};
    \addlegendentry{$\sigma=0$}
    \addplot table[x=n0, y=ave] {./output/subsample/macaque_0_01.csv};
    \addlegendentry{$\sigma=0.01$}
    \addplot table[x=n0, y=ave] {./output/subsample/macaque_0_05.csv};
    \addlegendentry{$\sigma=0.05$}
    \addplot table[x=n0, y=ave] {./output/subsample/macaque_0_10.csv};
    \addlegendentry{$\sigma=0.1$}
    \addplot table[x=n0, y=ave] {./output/subsample/macaque_0_20.csv};
    \addlegendentry{$\sigma=0.2$}
    
  \end{groupplot}
  
  \node[below right,fill=white] at (myplots c1r1.north west) {\huge\textbf{A}};
  \node[below right,fill=white] at (myplots c2r1.north west) {\huge\textbf{B}};
  \node[below right,fill=white] at (myplots c3r1.north west) {\huge\textbf{C}};
  \node[below right,fill=white] at (myplots c4r1.north west) {\huge\textbf{D}};
  
\end{tikzpicture}

}
    \vspace{-3mm}
  \caption{
    \textbf{(A)} Convergence of the inferred Laplacian for increasing graph size. `No $d(x)$' indicates the absence of a spectral shrinkage penalty, as in~\cite[Algorithm~3]{zhu2020network}.
    \textbf{(B)} Laplacian inference using noisy spectral templates and a graphon degree prior.
    \textbf{(C)} Inference of a synthetic graph Laplacian from noisy spectral templates using subgraph degree statistics.
    \textbf{(D)} Denoising of the Rhesus Macaque cerebral cortex network using subgraph degree statistics.
    \vspace{-0.5cm}
  }
  \label{fig:exp}
\end{figure*}

\section{Experiments}\label{sec:exp}

We demonstrate the utility of incorporating prior information on the graph model when inferring both real and synthetic networks.
Since we are interested in recovering the network structure and not the scale of $\mathbf{L}$, we normalize all graph matrices to have unit Frobenius norm before computing the error.
That is, for a ground-truth Laplacian $\L$ and an estimated Laplacian $\L^*$, our error function is
\begin{equation}\label{eq:exp-error-function}
  \mathrm{err}\left(\L^*,\L\right)=\left\|\frac{\L^*}{\|\L^*\|_F}-\frac{\L}{\|\L\|_F}\right\|_F.
\end{equation}
To minimize confounders when comparing methods, in estimating $\mathbf{L}^*$ we do not include any thresholding or rounding operation.
In solving~\eqref{eq:recovery-program}, we use the distance $\rho(\mathbf{A}, \mathbf{B}) = \| \mathbf{A} - \mathbf{B}\|_\mathrm{F}$, and set $\eta=0$ for the first experiment where the eigenvectors are perfectly known and $\eta = 2$ for all other experiments.

\parheading{Convergence with graph size.}
As stated in~\cref{thm:graphon-concentration}, the Laplacian spectrum concentrates near the graphon degree distribution as the sampled graph size grows.
To illustrate this, we estimate synthetic graphs of increasing sizes $n \in \{20,40,60,80,100\}$ drawn from the graphon $W(x,y)=\frac{1}{2} \gamma (x^2 + y^2) + (1-\gamma)$, where $\gamma=0.7$.
Three optimization programs are compared for graph estimation:
i)~Solving~\eqref{eq:recovery-program} for $d(x) = d_W(x)$ computed based on the true graphon $W$,
ii)~Solving~\eqref{eq:recovery-program} for $d(x) = d_{W'}(x)$ computed based on the alternate graphon $W'(x,y)=\frac{1}{2} (1-\gamma) (x^2 + y^2) + \gamma$,
and 
iii)~Solving~\eqref{eq:recovery-program} for $\beta=0$, as proposed in~\cite{zhu2020network}.
For the first two methods, the weight $\beta$ is chosen for optimal estimation for each graph size.
As shown in~\letterref{fig:exp}{A} for all graph sizes, including a spectral prior achieves a consistently higher accuracy than using no spectral information, even for the imperfect case when $W'$ is considered.
Moreover, Laplacian inference with the true generating graphon degree distribution uniformly attains minimal error.

\parheading{Noisy spectral templates.}
We demonstrate the estimation of a graph of size $n=40$ drawn from the graphon $W(x,y)=\frac{1}{2} \gamma (x^2 + y^2) + (1-\gamma)$ when noisy spectral templates are available.
These noisy eigenvectors are obtained from sample covariance matrices computed based on a varying number of observed filtered graph signals (cf. discussion before Remark~\ref{R:alternative_problems}), where the filter is given by $\Filt(\L)=\I-\alpha\L$.
The same three methods from the previous experiment are compared in \letterref{fig:exp}{B}.
Spectral shrinkage toward the true graphon degree distribution greatly improves estimation compared to the alternate graphon degree distribution or no spectral shrinkage.
This effect becomes more conspicuous for larger number of observed signals, i.e., for the setting where the Laplacian eigenvectors are better estimated.

\parheading{Incorporating subgraph degree statistics.}
While the previous two experiments assume knowledge of the graphon model, this might not be readily available in practice.
However, if the graph was indeed drawn from a graphon, its subgraphs are distributed according to the graphon as well.
In particular, we use the degree distribution of an observed subgraph for~\eqref{eq:recovery-program}, as discussed at the end of~\cref{sec:nti:approach}.
To understand the performance of using empirical degree distributions in~\eqref{eq:recovery-program}, we do not constrain the recovered graph to match the observed subgraph, although this would be done in practice.
This can also be interpreted as a valid illustration of the case where the degree distribution is estimated from a \emph{different} graph (as opposed to a subgraph) drawn from the same (unknown) graphon.

\subparheading{Synthetic example.}
The results of the described procedure for graphs drawn from the graphon $W(x,y)=1-0.8\max(x,y)$ and of size $n=100$ are plotted in \letterref{fig:exp}{C}.
We obtain a set of noisy spectral templates using the same graph filter and increasing number of graph signals as in the previous experiment.
Then, degree functions are extracted for subgraph sizes $n_0\in\{10,30,50,70,90\}$, as well as the underlying graphon for comparison.
Clearly, when the size of the observed subgraph increases, the estimated degree profile improves, yielding better performance.
Moreover, as the number of sampled graph signals $m$ increases, the inferred network improves in quality as well, due to the more accurately estimated spectral templates.
The experiment also reveals that, in the presence of noisy eigenvectors, approximate knowledge of the degree distribution can be almost as valuable as complete access to $d_W(x)$. 
The marginal value of the latter becomes apparent for large number of available signals.

\subparheading{Sparsifying a noisy brain network.}
We consider the connectome of the Rhesus Macaque's cerebral cortex~\cite{nr,markov2014weighted}, where each node ($|\Nodes|=91$) is a region in the cortex and each edge ($|\Edges|=1401$) corresponds to an interareal pathway.
For $n_0\in\{9, 18, 45, 91\}$ nodes, we randomly select $n_0$ nodes without replacement, and then compute the empirical degree function $\widehat{d}(x)$ of the subgraph as before.
Then, we inject Gaussian noise of increasing variance into the adjacency matrix, and use the eigenvectors of the corresponding Laplacian as noisy spectral templates in~\eqref{eq:recovery-program}.

As shown in~\letterref{fig:exp}{D}, as the subgraph size $n_0$ increases, the relative error between the recovered and true Laplacians decreases.
This indicates the utility of the degree function for the shrinkage penalty, since a more accurate estimate of the degree distribution leads to better recovery performance.
Additionally, as the power of the injected noise increases, the recovery performance degrades.
This is intuitive, as a corrupted set of spectral templates will bound the performance of the recovery algorithm.

\vspace{-1mm}
\section{Conclusion and future work}\label{sec:conc}

We considered the network topology inference problem through the lens of a graphon modeling assumption.
We demonstrated how the spectral properties of graph Laplacians drawn from a graphon relate to the degree function of that graphon, and how this relationship can be used to recover the graph topology from so-called spectral templates.
The proposed method based on this property was then extended to the case where the graphon is not known, but degree information is available from either a subgraph or an auxiliary graph.

Future research avenues will incorporate more of the rich structures and statistical properties afforded by graphon-based models.
Specifically, rather than relying on a shrinkage penalty for the Laplacian spectrum, one can relate the spectral moments of the adjacency matrix to the homomorphism densities of cycles in a graphon~\cite{lovasz2012large}.
Additionally, we will seek to generalize the considered penalties based on conventional graphons to the empirically relevant class of sparse exchangeable graph models~\cite{borgs2017graphons}.
Finally, we are currently investigating the use of spectral priors for the inference of higher-order network models, where edges are not strictly pairwise, but can relate arbitrarily large sets of nodes.

\vfill\pagebreak

\bibliographystyle{IEEEbib}
\bibliography{refs}

\end{document}